\documentclass[12pt,a4paper]{article}
\makeatletter
\def\eqnarray{%
\stepcounter{equation}%
\let\@currentlabel=\theequation
\global\@eqnswtrue
\global\@eqcnt\z@
\tabskip\@centering
\let\\=\@eqncr
$$\halign to \displaywidth\bgroup\@eqnsel\hskip\@centering
$\displaystyle\tabskip\z@{##}$&\global\@eqcnt\@ne
\hfil$\displaystyle{{}##{}}$\hfil
&\global\@eqcnt\tw@$\displaystyle\tabskip\z@{##}$\hfil
\tabskip\@centering&\llap{##}\tabskip\z@\cr}
\makeatother
\newcommand{\ket}[1]{{\vert{#1}\rangle}}
\newcommand{\bra}[1]{{\langle{#1}\vert}}
\newcommand{\braket}[2]{{\langle{#1}\vert{#2}\rangle}}

\newcommand{\fukuso}{{\mathbf C}}

\begin{document}

\title{\sl Generalized Bell States and Quantum Teleportation}
\author{
  Kazuyuki FUJII
  \thanks{E-mail address : fujii@math.yokohama-cu.ac.jp}\  
  \thanks{Home-page : http://fujii.sci.yokohama-cu.ac.jp}\\
  Department of Mathematical Sciences\\
  Yokohama City University\\
  Yokohama 236-0027\\ 
  JAPAN
  }
\date{}
\maketitle\thispagestyle{empty}
%
%
%
%
\begin{abstract}
  We make a brief comment on measurement of quantum operators with 
  degenerate eigenstates and apply to quantum teleportation.  We also 
  try extending the quantum teleportation by Bennett et al [5] to more 
  general situation by making use of generalized Bell states.
\end{abstract}

\newpage

%
%
%
%

\section{Introduction}
Quantum Teleportation is one of main subjects in Quantum Information Theory. 
See for example \cite{LPS}, \cite{AS}, \cite{APe}, \cite{AH} or \cite{KF1} 
on quantum information theory.

\par \noindent
This concept was first proposed by Bennett et al \cite{BBCJPW} and has  
been studying actively.  
Alice send (teleport) a quantum state localized near her to Bob 
making use of some manipulations. The significant feature of this is to use 
the so--called Bell states (\cite{JB}, \cite{BMR}) which are maximally 
entangled.  

In the process Alice must measure some physical operator which eigenstates 
are just the Bell states to know the final state through the reduction of 
state. Here much attension should be requireed. If a physical operator which 
Alice will use has degenerate eigenstates, then we cannot in general get one  
of Bell states explicitely by the principle of Quantum Mechanics. 
This has been pointed out by \cite{GA}. 
Alice must choose a physical operator with simple eigenvalues which 
eigenstates are the Bell states. 

By the way the generalized Bell states were defined by \cite{DF} making use 
of generalized coherent states and have been calculated by \cite{KF2}. 
Therefore it is very natural to try applying our generalized Bell states to 
quantum teleportation. Such a generalization is not so difficult 
(see Sect. 3). This is our main result.

\vspace{1cm}
\section{Review on Quantum Teleportation}
We in this section revisit the quantum teleportation by Bennett et al 
\cite{BBCJPW}. 

The famous Bell states (\cite{JB}, \cite{BMR}) in the case of spin 
$\frac{1}{2}$ are : 
\begin{eqnarray}
  \label{eq:Bell-states}
 && \ket{\phi_1}_{12}=\frac{1}{\sqrt{2}}
      (\ket{0}_1\otimes \ket{0}_2 + \ket{1}_1\otimes \ket{1}_2), \nonumber \\
 &&\ket{\phi_2}_{12}=\frac{1}{\sqrt{2}}
      (\ket{0}_1\otimes \ket{0}_2 - \ket{1}_1\otimes \ket{1}_2), \nonumber \\
 &&\ket{\phi_3}_{12}=\frac{1}{\sqrt{2}}
      (\ket{0}_1\otimes \ket{1}_2 + \ket{1}_1\otimes \ket{0}_2),  \nonumber \\
 &&\ket{\phi_4}_{12}=\frac{1}{\sqrt{2}}
      (\ket{0}_1\otimes \ket{1}_2 - \ket{1}_1\otimes \ket{0}_2).
\end{eqnarray}
These are a basis in ${\fukuso}^2\otimes {\fukuso}^2$. Conversely we have 
\begin{eqnarray}
 \label{eq:convert-relations}
&&\ket{0}_1\otimes \ket{0}_2=\frac{1}{\sqrt{2}}
   (\ket{\phi_1}_{12}+\ket{\phi_2}_{12}),\quad 
\ket{1}_1\otimes \ket{1}_2=\frac{1}{\sqrt{2}}
   (\ket{\phi_1}_{12}-\ket{\phi_2}_{12}), \nonumber \\
&&\ket{0}_1\otimes \ket{1}_2=\frac{1}{\sqrt{2}}
   (\ket{\phi_3}_{12}+\ket{\phi_4}_{12}),\quad 
\ket{1}_1\otimes \ket{0}_2=\frac{1}{\sqrt{2}}
   (\ket{\phi_3}_{12}-\ket{\phi_4}_{12}).
\end{eqnarray}

The quantum teleportation by Bennett et al \cite{BBCJPW} is as follows : 
Alice and Bob share beforehand a system of two particles $A$ and $B$ and  
they can constitute the EPR pair 
\begin{equation}
 \frac{1}{\sqrt{2}}
  \left(\ket{0}_A\otimes \ket{1}_B - \ket{1}_A\otimes \ket{0}_B \right)\ .
\end{equation}
Next Alice want to send (transport) a state 
\begin{equation}
    \alpha\ket{0}_1+\beta\ket{1}_1 \quad \mbox{where} \quad 
       \alpha\ ,\ \beta\  \in\  \fukuso
\end{equation}
to Bob. For that she manipulates a system of three particles $1$, $A$ and $B$
as follows : 
\par \vspace{5mm} 
\begin{flushleft}
 \begin{Large}
{\bf Fundamental Formula I}\ (\cite{BBCJPW}) 
 \end{Large}
\end{flushleft}
\begin{eqnarray}
 \label{eq:fundamental-formula1}
 &&(\alpha\ket{0}_1+\beta\ket{1}_1)\otimes 
 \frac{1}{\sqrt{2}}
  \left(\ket{0}_A\otimes \ket{1}_B-\ket{1}_A\otimes \ket{0}_B \right)
   \nonumber \\
 &&= \quad  
    \frac{1}{2}\ket{\phi_1}_{1A}\otimes (\alpha\ket{1}_B-\beta\ket{0}_B) 
  \ \ + \frac{1}{2}\ket{\phi_2}_{1A}\otimes (\alpha\ket{1}_B+\beta\ket{0}_B)
     \nonumber \\
 &&\quad\ 
  + \frac{1}{2}\ket{\phi_3}_{1A}\otimes (-\alpha\ket{0}_B+\beta\ket{1}_B)
  + \frac{1}{2}\ket{\phi_4}_{1A}\otimes (-\alpha\ket{0}_B-\beta\ket{1}_B)
\end{eqnarray}

\par \noindent
The proof is very easy making use of (\ref{eq:convert-relations}). 

\par \noindent
The procedure of quantum teleportation go as follows : 
\begin{itemize}
\item[(i)]\ Alice measures a physical operator $\widehat{Q}$ related to two 
     particles  system $\{1,\ A\}$ which eigenstates are 
     (\ref{eq:Bell-states}) and after that she obtains a state 
     (one of (\ref{eq:Bell-states})) by the reduction of state. 
\vspace{2mm}
\item[(ii)]\ Alice informs this outcome to Bob by some classical means of 
     communication (there is no inconsistency to the theory of special 
     relativity).
\vspace{2mm}
\item[(iii)]\ Bob knows by this what the state corresponding to 
    a particle $B$ is  
\[
   \alpha\ket{1}_B-\beta\ket{0}_B,\quad 
   \alpha\ket{1}_B+\beta\ket{0}_B,\quad 
   -\alpha\ket{0}_B+\beta\ket{1}_B,\quad 
   -\alpha\ket{0}_B-\beta\ket{1}_B.
\]
%
\item[(iv)]\ Bob operates some operators to get the final result 
\begin{eqnarray}
 &&i\sigma_2(\alpha\ket{1}_B-\beta\ket{0}_B)=\alpha\ket{0}_B+\beta\ket{1}_B,
  \nonumber \\  
 &&\sigma_1(\alpha\ket{1}_B+\beta\ket{0}_B)=\alpha\ket{0}_B+\beta\ket{1}_B,
  \nonumber \\ 
 &&-\sigma_3(-\alpha\ket{0}_B+\beta\ket{1}_B)=\alpha\ket{0}_B+\beta\ket{1}_B,
  \nonumber \\  
 &&-{\bf 1}_2(-\alpha\ket{0}_B-\beta\ket{1}_B)=\alpha\ket{0}_B+\beta\ket{1}_B.
\end{eqnarray}
\end{itemize}

\par \noindent
Here we have something to worry. In (i) Alice measures a physical operator 
$\widehat{Q}$. 
\begin{flushleft}
 \begin{Large}
{\bf What a kind of physical operator does she measure ?}
 \end{Large}
\end{flushleft}
If we measure physical operators like 
\begin{equation}
 \label{eq:operator12}
   \widehat{Q}=\sigma_1\otimes \sigma_1 + \sigma_2\otimes \sigma_2
\end{equation}
or
\begin{equation}
 \label{eq:operator13}
   \widehat{Q}=\sigma_1\otimes \sigma_1 + \sigma_3\otimes \sigma_3
\end{equation}
then we meet some troubles. This has been pointed out by Adenier \cite{GA}
\footnote{I don't agree to his assertion in spite of his good pointout}. 
For example let us consider the operator (\ref{eq:operator13}) which 
eigenvalues are $\{2, 0,  -2\}$ and corresponding eigenstates are 
$\ket{\phi_1}, \{\ket{\phi_2}, \ket{\phi_3}\}, \ket{\phi_1}$  in this order. 
Namely $0$ is an eigenvalue with multiplicity $2$.

\par \noindent
Let us here assume that we get $0$--eigenvalue when measuring $\widehat{Q}$.  
What is the state we get after the reduction of state ? \quad  Since 
\[
    \ket{0} \in \mbox{Vect}_{\fukuso}\{\ket{\phi_2},\ket{\phi_3}\}, 
\]
we have some 
\begin{equation}
    \alpha\ket{\phi_2}+\beta\ket{\phi_3}\quad \mbox{where}\quad 
           \alpha,\ \  \beta\ \ \in \ \ \fukuso
\end{equation}
not
\begin{equation}
     \ket{\phi_2}\quad \mbox{or}\quad \ket{\phi_3}
\end{equation}
by the principle of Quantum Mechanics.  
That is, it is dangerous for us to use a physical operator with 
degenerate eigenstates in the process of measurement. 

\par \noindent 
Therfore we must use a physical operator $\widehat{Q}$ with simple 
eigenvalues which corresponding eigenstates are just (\ref{eq:Bell-states}). 
Let us consider an operator 
\begin{equation}
 \label{eq:physical-operator1}
   \widehat{Q}=a\ket{\phi_1}\bra{\phi_1}+b\ket{\phi_2}\bra{\phi_2}
     +c\ket{\phi_3}\bra{\phi_3}+d\ket{\phi_4}\bra{\phi_4}
\end{equation}
where $a, b, c, d$ are mutually distinct real numbers. Then it is clear 
that the eigenvalues of $\widehat{Q}$ are $\{a,b,c,d\}$ and 
corresponding eigenstates are $\{\ket{\phi_1},\ket{\phi_2},\ket{\phi_3},
\ket{\phi_4}\}$ in this order. 

\par \noindent 
We want to rewrite (\ref{eq:physical-operator1}) making use of 
\begin{eqnarray}
  &&\ket{0}\bra{0}=\frac{1}{2}({\bf 1}_2+\sigma_3),\quad \ 
    \ket{1}\bra{1}=\frac{1}{2}({\bf 1}_2-\sigma_3), \nonumber \\ 
  &&\ket{0}\bra{1}=\frac{1}{2}(\sigma_1+i\sigma_2),\quad 
    \ket{1}\bra{0}=\frac{1}{2}(\sigma_1-i\sigma_2).
\end{eqnarray}
The result reads  
\begin{eqnarray}
  \widehat{Q}=
    &&\frac{a+b+c+d}{4}{\bf 1}_2\otimes {\bf 1}_2  + 
    \frac{a-b+c-d}{4}\sigma_1\otimes \sigma_1   \nonumber \\
  +
    &&\frac{-a+b+c-d}{4}\sigma_2\otimes \sigma_2  +
    \frac{a+b-c-d}{4}\sigma_3\otimes \sigma_3 \ . 
\end{eqnarray}
The proof is easy.  

\par \noindent 
For simplicity we set $a=3,\ b=1,\ c=-1,\ d=-3$ to get a simple form   
\begin{equation}
   \widehat{Q}=\sigma_1\otimes \sigma_1 + 2\sigma_3\otimes \sigma_3\ .
\end{equation}
That is, Alice should measure this operator in place of (\ref{eq:operator13}) 
or (\ref{eq:operator12}).

\vspace{2cm}
\section{More on Quantum Teleportation}
We extend the quantum teleportation by Bennett et al \cite{BBCJPW} to 
more general situation. In the following  
we treat the coherent representation of $u(3)$ based on 
$\frac{U(3)}{U(2)\times U(1)} \cong {\fukuso}P^2$ (see \cite{KF2} and 
\cite{FKSF}) with $Q=1$ only to avoid complicated situations. 
This is just the three dimensional representation. 

\par \noindent
Let $\{\ket{0},\ket{1},\ket{2}\}$ be a basis of the representation space 
$V$ ($\cong {\fukuso}^3$). Namely 
\[
 \sum_{j=0}^{2}\ket{j}\bra{j}={\bf 1}_{Q=1}
 \quad \mbox{and} \quad  \braket{i}{j}=\delta_{ij}.
\]

\par \noindent
Let $\omega$ be an element in $\fukuso$ satisfying ${\omega}^3=1$. 
Then $1+\omega+{\omega}^2=0$ and ${\bar \omega}={\omega}^2$. 
Then generalized Bell states in this case ($n=2$ and $Q=1$) are given by 
\cite{KF2} : 
\begin{eqnarray}
 \label{eq:gBell-states}
  &&\ket{\psi_1}_{12}=\frac{1}{\sqrt{3}}
  (\ket{0}_1\otimes \ket{0}_2+\ket{1}_1\otimes \ket{1}_2
     +\ket{2}_1\otimes \ket{2}_2), \nonumber \\
  &&\ket{\psi_2}_{12}=\frac{1}{\sqrt{3}}
  (\ket{0}_1\otimes \ket{0}_2+\omega\ket{1}_1\otimes \ket{1}_2+
   \omega^2\ket{2}_1\otimes \ket{2}_2), \nonumber \\
  &&\ket{\psi_3}_{12}=\frac{1}{\sqrt{3}}
  (\ket{0}_1\otimes \ket{0}_2+\omega^2\ket{1}_1\otimes \ket{1}_2+
   \omega\ket{2}_1\otimes \ket{2}_2), \nonumber \\
  &&\ket{\psi_4}_{12}=\frac{1}{\sqrt{3}}
  (\ket{0}_1\otimes \ket{1}_2+\ket{1}_1\otimes \ket{2}_2+
     \ket{2}_1\otimes \ket{0}_2), 
  \nonumber \\
  &&\ket{\psi_5}_{12}=\frac{1}{\sqrt{3}}
  (\ket{0}_1\otimes \ket{1}_2+\omega\ket{1}_1\otimes \ket{2}_2+
   \omega^2\ket{2}_1\otimes \ket{0}_2), \nonumber \\
  &&\ket{\psi_6}_{12}=\frac{1}{\sqrt{3}}
  (\ket{0}_1\otimes \ket{1}_2+\omega^2\ket{1}_1\otimes \ket{2}_2+
   \omega\ket{2}_1\otimes \ket{0}_2), \nonumber \\
  &&\ket{\psi_7}_{12}=\frac{1}{\sqrt{3}} 
  (\ket{0}_1\otimes \ket{2}_2+\ket{1}_1\otimes \ket{0}_2+
    \ket{2}_1\otimes \ket{1}_2), 
  \nonumber \\
  &&\ket{\psi_8}_{12}=\frac{1}{\sqrt{3}}
  (\ket{0}_1\otimes \ket{2}_2+\omega\ket{1}_1\otimes \ket{0}_2+
   \omega^2\ket{2}_1\otimes \ket{1}_2), \nonumber \\
  &&\ket{\psi_9}_{12}=\frac{1}{\sqrt{3}}
  (\ket{0}_1\otimes \ket{2}_2+\omega^2\ket{1}_1\otimes \ket{0}_2+
   \omega\ket{2}_1\otimes \ket{1}_2).
\end{eqnarray}
These are a basis in $V\otimes V\  \cong\  {\fukuso}^3\otimes {\fukuso}^3$. 
Conversely we have 
\begin{eqnarray}
 \label{eq:convert-formula2}
 &&\ket{0}_1\otimes \ket{0}_2 =\frac{1}{\sqrt{3}}
 (\ket{\psi_1}_{12}+\ket{\psi_2}_{12}+\ket{\psi_3}_{12}), \nonumber \\
 &&\ket{1}_1\otimes \ket{1}_2 =\frac{1}{\sqrt{3}}
 (\ket{\psi_1}_{12}+\omega^2\ket{\psi_2}_{12}+\omega\ket{\psi_3}_{12}),
  \nonumber \\
 &&\ket{2}_1\otimes \ket{2}_2 =\frac{1}{\sqrt{3}}
 (\ket{\psi_1}_{12}+\omega\ket{\psi_2}_{12}+\omega^2\ket{\psi_3}_{12}),
  \nonumber \\
 &&\ket{0}_1\otimes \ket{1}_2 =\frac{1}{\sqrt{3}}
 (\ket{\psi_4}_{12}+\ket{\psi_5}_{12}+\ket{\psi_6}_{12}),  \nonumber \\
 &&\ket{1}_1\otimes \ket{2}_2 =\frac{1}{\sqrt{3}}
 (\ket{\psi_4}_{12}+\omega^2\ket{\psi_5}_{12}+\omega\ket{\psi_6}_{12}),
  \nonumber \\
 &&\ket{2}_1\otimes \ket{0}_2 =\frac{1}{\sqrt{3}}
 (\ket{\psi_4}_{12}+\omega\ket{\psi_5}_{12}+\omega^2\ket{\psi_6}_{12}),
  \nonumber \\
 &&\ket{0}_1\otimes \ket{2}_2 =\frac{1}{\sqrt{3}}
 (\ket{\psi_7}_{12}+\ket{\psi_8}_{12}+\ket{\psi_9}_{12}), \nonumber \\
 &&\ket{1}_1\otimes \ket{0}_2 =\frac{1}{\sqrt{3}}
 (\ket{\psi_7}_{12}+\omega^2\ket{\psi_8}_{12}+\omega\ket{\psi_9}_{12}),
  \nonumber \\
 &&\ket{2}_1\otimes \ket{1}_2 =\frac{1}{\sqrt{3}}
 (\ket{\psi_7}_{12}+\omega\ket{\psi_8}_{12}+\omega^2\ket{\psi_9}_{12}).
\end{eqnarray}

\par \noindent
Alice and Bob share beforehand a system of two particles $A$ and $B$ and  
they can constitute the EPR--like pair  
\begin{equation}
 \frac{1}{\sqrt{3}}
  \left(\ket{0}_A\otimes \ket{1}_B + 
        \omega\ket{1}_A\otimes \ket{2}_B + 
        \omega^2\ket{2}_A\otimes \ket{0}_B 
  \right)\ .
\end{equation}
Next Alice want to send (transport) a state 
\begin{equation}
    \alpha\ket{0}_1+\beta\ket{1}_1+\gamma\ket{2}_1
  \quad \mbox{where} \quad 
       \alpha\ ,\ \beta\ ,\ \gamma\ \in\  \fukuso
\end{equation}
to Bob. For that she manipulates a system of three particles $1$, $A$ and $B$
as follows : 
\par \vspace{5mm}
\begin{flushleft}
 \begin{Large}
{\bf Fundamental Formula II}
 \end{Large}
\end{flushleft}
\begin{eqnarray}
 \label{eq:fundamental-formula2}
 &&(\alpha\ket{0}_1+\beta\ket{1}_1+\gamma\ket{2}_1)\otimes 
  \frac{1}{\sqrt{3}}
    \left(\ket{0}_A\otimes \ket{1}_B + 
        \omega\ket{1}_A\otimes \ket{2}_B +
        \omega^2\ket{2}_A\otimes \ket{0}_B 
  \right)   \nonumber \\
 &&= \frac{1}{3}\{\ \ 
     \ket{\psi_1}_{1A}\otimes \left(
     \alpha\ket{1}_B+\beta\omega\ket{2}_B+\gamma\omega^2\ket{0}_B \right)
             \nonumber \\
  &&\qquad +\ \ket{\psi_2}_{1A}\otimes \left(
     \alpha\ket{1}_B+\beta\ket{2}_B+\gamma\ket{0}_B \right) 
           \nonumber \\
  &&\qquad +\ \ket{\psi_3}_{1A}\otimes \left(
     \alpha\ket{1}_B+\beta\omega^2\ket{2}_B+\gamma\omega\ket{0}_B \right) 
           \nonumber \\
  &&\qquad +\ \ket{\psi_4}_{1A}\otimes \left(
     \alpha\omega\ket{2}_B+\beta\omega^2\ket{0}_B+\gamma\ket{1}_B \right) 
           \nonumber \\
  &&\qquad +\ \ket{\psi_5}_{1A}\otimes \left(
     \alpha\omega\ket{2}_B+\beta\omega\ket{0}_B+\gamma\omega\ket{1}_B \right) 
           \nonumber \\
  &&\qquad +\ \ket{\psi_6}_{1A}\otimes \left(
     \alpha\omega\ket{2}_B+\beta\ket{0}_B+\gamma\omega^2\ket{1}_B \right) 
           \nonumber \\
  &&\qquad +\ \ket{\psi_7}_{1A}\otimes \left(
     \alpha\omega^2\ket{2}_B+\beta\ket{0}_B+\gamma\omega\ket{1}_B \right) 
           \nonumber \\
  &&\qquad +\ \ket{\psi_8}_{1A}\otimes \left(
     \alpha\omega^2\ket{2}_B+\beta\omega^2\ket{0}_B+\gamma\omega^2\ket{1}_B 
   \right)   \nonumber \\
  &&\qquad +\ \ket{\psi_9}_{1A}\otimes \left(
     \alpha\omega^2\ket{2}_B+\beta\omega\ket{0}_B+\gamma\ket{1}_B \right)\ \  
\} .
\end{eqnarray}
The proof is straightforward. First expand the left hand side of 
(\ref{eq:fundamental-formula2}) and next rearrange each terms making use of 
(\ref{eq:convert-formula2}). We leave details to the readers.

\par \noindent
As shown in the preceeding section Alice must measure a physical operator 
$\widehat{Q}$ with non--degenerate eigenstates $\{\ket{\psi_1}, \ket{\psi_2}, 
\cdots, \ket{\psi_9}\}$ such as 
\begin{equation}
 \label{eq:physical-operator2}
  \widehat{Q}=\sum_{j=1}^{9} a_j \ket{\psi_j}\bra{\psi_j}
\end{equation}
with mutually distinct real numbers $\{a_1, a_2, \cdots, a_9\}$ . 
We want to rewrite (\ref{eq:physical-operator2}) making use of 
\begin{eqnarray}
 \label{eq:xy}
 &&Z=\ket{0}\bra{0}+\omega\ket{1}\bra{1}+\omega^2\ket{2}\bra{2},\quad \  
   X=\ket{1}\bra{0}+\ket{2}\bra{1}+\ket{0}\bra{2}
 \nonumber \\
 &&Z^2=\ket{0}\bra{0}+\omega^2\ket{1}\bra{1}+\omega\ket{2}\bra{2},\quad  
   X^2=\ket{2}\bra{0}+\ket{0}\bra{1}+\ket{1}\bra{2} 
 \nonumber \\
 &&Z^3={\bf 1}_3, \qquad \qquad \qquad \qquad \qquad \quad \ 
   X^3={\bf 1}_3\ .
\end{eqnarray}
For example 
\begin{eqnarray}
 \label{eq:several-xy}
  &&\ket{0}\bra{0}=\frac{1}{3}\left({\bf 1}_3 + Z + Z^2 \right), 
      \nonumber \\
  &&\ket{1}\bra{1}=\frac{1}{3}\left({\bf 1}_3 + XZX^2 + XZ^2X^2 \right),
      \nonumber \\ 
  &&\ket{2}\bra{2}=\frac{1}{3}\left({\bf 1}_3 + X^2ZX + X^2Z^2X \right),
      \nonumber \\ 
  &&\ket{0}\bra{1}=\frac{1}{3}\left(X^2 + ZX^2 + Z^2X^2 \right),
      \nonumber \\ 
  &&\ket{1}\bra{0}=\frac{1}{3}\left(X + XZ + XZ^2 \right),
      \nonumber \\ 
  &&\ket{0}\bra{2}=\frac{1}{3}\left(X + ZX + Z^2X \right),
      \nonumber \\ 
  &&\ket{2}\bra{0}=\frac{1}{3}\left(X^2 + X^2Z + X^2Z^2 \right),
      \nonumber \\ 
  &&\ket{1}\bra{2}=\frac{1}{3}\left(X^2 + XZX + XZ^2X \right),
      \nonumber \\ 
  &&\ket{2}\bra{1}=\frac{1}{3}\left(X + X^2ZX^2 + X^2Z^2X^2 \right).
\end{eqnarray}
\par \vspace{5mm} \noindent
We can with certainty rewrite (\ref{eq:physical-operator2}) makig use of 
(\ref{eq:xy}) and (\ref{eq:several-xy}) though the calculations are 
miserable. We leave it to the readers.

After receiving an information by Alice Bob operates some operators to get 
the final result
\begin{eqnarray}
  &&Z^2X^2 \left(
     \alpha\ket{1}_B+\beta\omega\ket{2}_B+\gamma\omega^2\ket{0}_B 
      \right)
    = \alpha\ket{0}_B+\beta\ket{1}_B+\gamma\ket{2}_B ,
  \nonumber \\
  &&X^2 \left(
     \alpha\ket{1}_B+\beta\ket{2}_B+\gamma\ket{0}_B 
      \right)
    = \alpha\ket{0}_B+\beta\ket{1}_B+\gamma\ket{2}_B ,
  \nonumber \\
  &&ZX^2 \left(
     \alpha\ket{1}_B+\beta\omega^2\ket{2}_B+\gamma\omega\ket{0}_B 
      \right)
    = \alpha\ket{0}_B+\beta\ket{1}_B+\gamma\ket{2}_B ,
  \nonumber \\
  &&\omega^2Z^2X \left(
     \alpha\omega\ket{2}_B+\beta\omega^2\ket{0}_B+\gamma\ket{1}_B 
      \right)
    = \alpha\ket{0}_B+\beta\ket{1}_B+\gamma\ket{2}_B ,
  \nonumber \\
  &&\omega^2X \left(
     \alpha\omega\ket{2}_B+\beta\omega\ket{0}_B+\gamma\omega\ket{1}_B 
      \right)
    = \alpha\ket{0}_B+\beta\ket{1}_B+\gamma\ket{2}_B ,
  \nonumber \\
  &&\omega^2ZX \left(
     \alpha\omega\ket{2}_B+\beta\ket{0}_B+\gamma\omega^2\ket{1}_B 
      \right)
    = \alpha\ket{0}_B+\beta\ket{1}_B+\gamma\ket{2}_B ,
  \nonumber \\
  &&\omega Z^2 \left(
     \alpha\omega^2\ket{0}_B+\beta\ket{1}_B+\gamma\omega\ket{2}_B 
      \right)
    = \alpha\ket{0}_B+\beta\ket{1}_B+\gamma\ket{2}_B ,
  \nonumber \\
  &&\omega{\bf 1}_3 \left(
     \alpha\omega^2\ket{0}_B+\beta\omega^2\ket{1}_B+\gamma\omega^2\ket{2}_B 
      \right)
    = \alpha\ket{0}_B+\beta\ket{1}_B+\gamma\ket{2}_B ,
  \nonumber \\
  &&\omega Z \left(
     \alpha\omega^2\ket{0}_B+\beta\omega\ket{1}_B+\gamma\ket{2}_B 
      \right)
    = \alpha\ket{0}_B+\beta\ket{1}_B+\gamma\ket{2}_B .
\end{eqnarray}
This is an extended version of quantum teleportation by Bennett et al 
\cite{BBCJPW}.

\vspace{1cm}
\section{Discussion}
In this paper we discussed the importance to use a physical operator with 
non--degenerate eigenstates (the Bell states) in the process of 
measurement and next applied our generalized Bell states to quantum 
teleportation. 

\par \noindent
The generalized Bell states including usual Bell states are deeply related 
to a (compact) complex geometry  \cite{KF2}, \cite{KF1}, \cite{FKS},\   
so quantum teleportation should be in our opinion more geometrized. 

\par \noindent
We want to give a unfied geometric approach to Quantum Information 
Theory including quantum computer, quantum teleportation, 
quantum cryptgraphy.  

\par \noindent
In \cite{KF3} such a trial will be given.

\vspace{2cm}
\noindent
{\it Acknowledgment.}
The author wishes to thank Shin'ichi Nojiri for useful discussions with him  
and Asher Peres for pointing out a misunderstanding on the Bell states. 
%


\end{document}